\documentclass[runningheads]{llncs}
\usepackage{makeidx}
\usepackage[utf8]{inputenc}
\usepackage{enumitem} % for (i), (ii), ... in enumerate environments
\usepackage{listings} % code listings
\usepackage{color} % default colors
\usepackage{hyperref} % hyperlinks
\usepackage{graphicx} % illustrations
\usepackage{multirow} % multirow tables
\definecolor{dkgreen}{rgb}{0,0.6,0}
\definecolor{gray}{rgb}{0.5,0.5,0.5}
\definecolor{mauve}{rgb}{0.58,0,0.82}
\definecolor{lightgray}{rgb}{0.8,0.8,0.8}
\definecolor{darkblue}{rgb}{0,0,.5}
\hypersetup{
	colorlinks=true,
	breaklinks=true,
	linkcolor=darkblue,
	menucolor=darkblue,
	urlcolor=darkblue,
	citecolor=darkblue,
	pagecolor=darkblue,
	filecolor=darkblue
}
\begin{document}
\title{{\small\textnormal{Technical Report TR-2014-1}}\\\vspace{18pt}
GR2RSS: Publishing Linked Open Commerce Data as RSS and Atom Feeds}
\titlerunning{Publishing Linked Open Commerce Data as RSS and Atom Feeds}
\author{Alex Stolz \and Martin Hepp}
\authorrunning{A. Stolz and M. Hepp}
\institute{
	E-Business and Web Science Research Group, Universität der Bundeswehr München\\
	Werner-Heisenberg-Weg 39, D-85579 Neubiberg, Germany\\
	\email{\{alex.stolz,martin.hepp\}@unibw.de}
}
\maketitle

\pagenumbering{arabic}

\begin{abstract}
The integration of Linked Open Data (LOD) content in Web pages is a challenging and sometimes tedious task for Web developers. At the same moment, most software packages for blogs, content management systems~(CMS), and shop applications support the consumption of feed formats, namely RSS and Atom.
In this technical report, we demonstrate an on-line tool that fetches e-commerce data from a SPARQL endpoint and syndicates obtained results as RSS or Atom feeds. Our approach combines
(1)~the popularity and broad tooling support of existing feed formats,
(2)~the precision of queries against structured data built upon common Web vocabularies like schema.org, GoodRelations, FOAF, VCard, and WGS~84, and
(3)~the ease of integrating content from a large number of Web sites and other data sources in RDF in general.
%We show that our Semantic Web based approach is competitive with similar existing tools.
\end{abstract}

\section{Introduction}

Despite the growing amount of structured data on the Web, the useful integration into Web pages still lags behind opportunities. In the field of e-commerce, major retail sites like \textit{sears.com}, \textit{bestbuy.com}, \textit{wayfair.com}, \textit{rakuten.de} and numerous smaller shops have added RDFa and Microdata markup to their page templates, exposing more than 30 million offers that are updated on a daily basis.
% http://wiki.goodrelations-vocabulary.org/Datasets
A couple of SPARQL endpoints already collate this structured e-commerce data. However, fetching useful information typically involves contacting the right endpoints, crafting proper SPARQL queries, and eventually converting results into data formats understood by target applications. Average Web developers and site owners quickly get overwhelmed by the technical challenges imposed by these tasks.
At the same time, there exist popular data formats for publishing dynamic content on the Web, namely RSS~\cite{rss2.0} and the Atom Syndication Format~\cite{Nottingham2005,Sayre2005}.
%Such feed formats provide incentives for both publishers and subscribers: Publishers can increase the reach of their content, and that often in very qualified audiences. Subscribers can augment their pages by fresh, quality content that they could not produce themselves.
Both feed formats have excellent tool support; i.e., most software packages for blogs, content management systems (CMS), and shopping carts provide integration capabilities for external sources using RSS or Atom.

In this technical report, we show an approach that combines the broad tooling support of established data feed standards with the precision of queries against structured data collected from multiple Web sites and RDF data sources, built upon common Web vocabularies such as schema.org\footnote{\url{http://schema.org/}}, GoodRelations~\cite{Hepp2008}, FOAF\footnote{\url{http://xmlns.com/foaf/spec/}}, VCard\footnote{\url{http://www.w3.org/TR/vcard-rdf/}}, and WGS~84\footnote{\url{http://www.w3.org/2003/01/geo/}}.
%Our on-line tool comprises a set of query builders, a caching mechanism, and a currency conversion at the endpoint. Moreover, it implements GeoRSS information and embeds RDFa into feed formats.

\section{GR2RSS Tool}

%Given the popularity of RSS and Atom and the ease of integrating content from a large number of Web sites and other RDF data sources, we propose an on-line tool that permits to pose queries against SPARQL endpoints.
%Publicly available endpoints that provide a consolidated view on a large share of commerce data on the Web are LOC\footnote{\url{http://linkedopencommerce.com}}, LOD\footnote{\url{http://lod.openlinksw.com/}} and URIBurner\footnote{\url{http://uriburner.com/}}.

\begin{figure}[ht]
	\centering
	\includegraphics[width=0.7\textwidth]{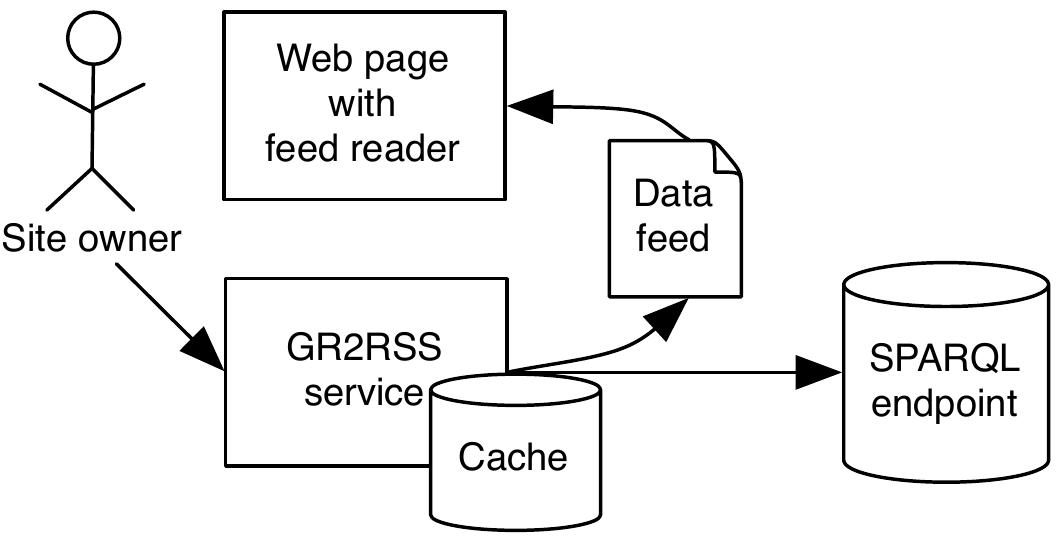}
	\caption{Conceptual architecture of the GR2RSS on-line service}
	\label{fig1}
\end{figure}

The on-line tool\footnote{\url{http://www.ebusiness-unibw.org/tools/gr2rss/}} that we have developed fetches GoodRelations e-commerce data from a SPARQL endpoint and syndicates obtained results as RSS or Atom feeds. Fig. \ref{fig1} outlines the general system architecture of our tool. The generated data feeds serve as carriers to facilitate the consumption and integration of linked open commerce data in Web pages. This way site owners have a means to add relevant, dynamic content to their Web pages, thereby attracting visitors and improving search engine rankings. Conversely, product vendors gain additional visibility of their products for free, since items republished by virtue of site owners and bloggers link back to shops where the products are actually offered.

%Our service aims to provide enhanced flexibility and usability. It allows to search for products or physical stores of legal entities.
%Optional query builders assist users in creating respective queries. After the successful execution of a query, two links are presented to the user that correspond to the generated data feeds. The URI patterns of the feed links do not reveal the actual SPARQL query. Instead, the query is encoded behind an incremental numeric identifier that maps to a local database record which stores the SPARQL query. This way URIs are kept short and readable and, at the same time, a potentially unintentional disclosure of SPARQL expert knowledge is prevented.

In the following, we summarize the core technical contributions of our on-line service:
%Other important features that our on-line tool provides include a universal caching mechanism in order to alleviate high traffic at the endpoint for recurring queries. Geo location data for displaying feeds in maps is incorporated, repetitive RDFa annotation of results, and live currency conversion in the endpoint.

\begin{enumerate}
\item \textit{Query builder:}
The tool supports three different levels of expertise. Depending on the selected search mode, a more or less sophisticated query builder is presented to the user. The following three search modes are available:
\begin{itemize}
	\item \textit{Basic:} Single input field for keyword searches.
	\item \textit{Extended:} Query builder with support for
		filtering and sorting results,
		price currency conversion for products, and
		location-aware store searches.
	\item \textit{Expert:} Input field for entering raw SPARQL queries and a set of possible variables which bindings can be processed by the feed templates.
\end{itemize}

\item \textit{Prefetching and caching:}
In order to limit load at the SPARQL endpoints, we implemented server-side caching of the generated feeds.
Our caching mechanism stores records of successfully executed queries in a local database management system for serving future requests and creates corresponding cache files. After a given cache period (e.g. a day), future requests trigger cache invalidations whereby the cached files are replaced by freshly generated feed content.
%Our caching mechanism is universal and independent of SPARQL endpoints or protocols used (cf. ).

\item \textit{Geo information:}
RSS and Atom feed formats provide extension mechanisms to support custom vocabularies. We used the GeoRSS\footnote{\url{http://georss.org/}} vocabulary to include positional data in data feeds. These annotations allow, for instance, to extract and display location data on a map, e.g. on \textit{Google Maps}.

\item \textit{Viral use of RDFa:}
%Numerous blog systems and CMS can tap RSS and Atom feeds.
Based on the idea of embedding RSS and Atom feeds in blog systems and CMS, we decided to piggyback RDFa statements as entity-encoded HTML in feed entries \cite{Stolz2013:Hypertext}. Thereby we can obtain a viral publication effect, because RDFa preserves the URIs of the original entities and thus prevents the proliferation of identifiers. In other words, any Web page integration of feed content contributes to the promotion of the product offers at the origin.
Moreover, we employ \textit{foaf:page} links to provide a means for tracking the document URI at which the particular content reappears.
%gradually attaching it to the graph of e-commerce data.

\item \textit{Currency conversion:}
The currency conversion at the endpoint is realized using a materialization of exchange rates based on the Exchange Rate Ontology\footnote{\url{http://purl.org/xro/}} (XRO). These currency exchange rates need to be available before any currency conversion task in SPARQL can take place. A service based on XRO that provides regularly updated exchange rates in RDF is available at \url{http://www.currency2currency.org/} \cite{Stolz2013:SALAD}.
% The graph pattern in Listing~\ref{lst:sparql_query} exemplifies how a list of product prices in the range of 10 and 20 euros can be determined based on GoodRelations and XRO:
%\begin{lstlisting}[language=SQL, morekeywords={OPTIONAL, FILTER, contains, a}, caption={Graph pattern of a SPARQL query that normalizes to euros}, label=lst:sparql_query]
%?pspec a gr:UnitPriceSpecification .
%?pspec gr:hasCurrency ?currency .
%?pspec gr:hasCurrencyValue ?pval .
%?to_curr xro:hasCode ?to_code .
%FILTER(contains(?to_code, "EUR"))
%?to_curr xro:hasRate ?to_rate .
%?from_curr xro:hasCode ?currency .
%?from_curr xro:hasRate ?from_rate .
%FILTER(?pval*?to_rate/?from_rate>=10 && ?pval*?to_rate/?from_rate<=20)
%\end{lstlisting}

\end{enumerate}

%% URIBurner (works!):
%PREFIX gr: <http://purl.org/goodrelations/v1#>
%PREFIX xro: <http://purl.org/xro/ns#>
%
%SELECT DISTINCT ?offer, ?name, ?currency, ?from_rate, ?pval as ?from_pval, ?code, ?to_rate, ?pval*?to_rate/?from_rate as ?to_pval
%WHERE {
%  ?offer a gr:Offering .
%  ?offer gr:name ?name .
%  ?offer gr:hasPriceSpecification ?pspec .
%  ?pspec a gr:UnitPriceSpecification .
%  ?pspec gr:hasCurrency ?currency .
%  ?pspec gr:hasCurrencyValue ?pval .
%  ?to_curr xro:hasCode ?code .
%  FILTER(contains(?code, "EUR"))
%  ?to_curr xro:hasRate ?to_rate .
%  ?from_curr xro:hasCode ?currency .
%  ?from_curr xro:hasRate ?from_rate .
%  FILTER(isNumeric(?pval))
%  FILTER(?pval*?to_rate/?from_rate>=10 && ?pval*?to_rate/?from_rate<=20)
%}
%LIMIT 10

Our on-line tool was written in PHP and uses Javascript for user interaction. It runs over Linked Open Commerce data stores, and is compatible with Virtuoso SPARQL endpoints that support the \emph{bif:contains} feature, a built-in function that executes over a full-text search index. For the future, we are planning to rewrite parts of the code in order to make the tool SPARQL-1.1-compliant.

%Currently, the distances computation feature between geo positions is disabled for the open source edition of Virtuoso, thus, queries that rely on geo functionality will fail on such instances. 
%In addition, the currency conversion functionality used for advanced querying is using a small knowledge base for currency exchange rates. If the latter has not been loaded properly into the endpoint, then no results can be retrieved when the currency conversion option is set.

\section{Demonstration}
\begin{figure}[ht]
	\centering
	\includegraphics[width=1.0\textwidth]{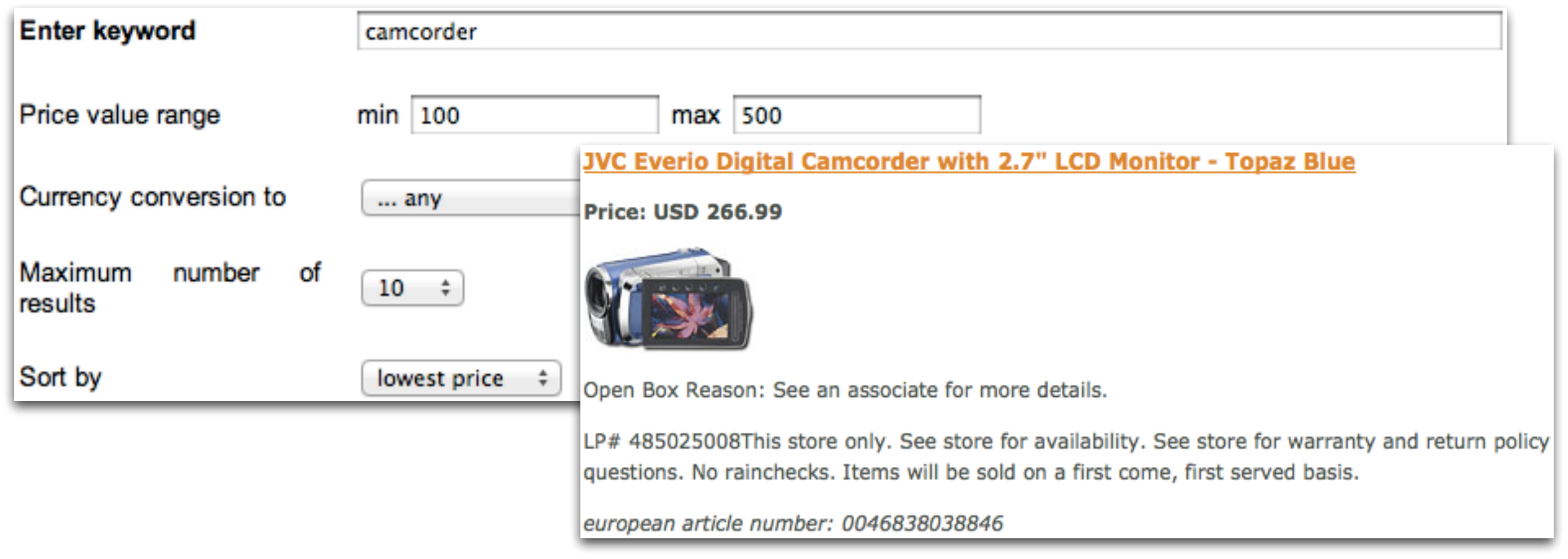}
	\caption{Query builder for \textit{camcorder} search and Web page integration of the feed}
	\label{fig2}
\end{figure}

In the following, we demonstrate an example of integrating a generated feed into a Web page.
Suppose that we are looking for products within a price range of $100$ and $500$ dollars that contain the keyword ``camcorder'' and include product pictures.
The populated form fields of the query builder are depicted in Fig.~\ref{fig2}.
The figure also shows a single camcorder item after integrating the feed into a Web page.
Similar examples can be found in the tool documentation\footnote{\url{http://www.ebusiness-unibw.org/tools/gr2rss/docs}} and example page\footnote{\url{http://www.ebusiness-unibw.org/tools/gr2rss/examples}}.

\section{Related Work}

We compared our work with existing approaches, namely (1) single feed definition dialogs, as offered by major sites (e.g. eBay and Amazon), and (2) feed aggregation services, i.e. Yahoo Pipes\footnote{\url{http://pipes.yahoo.com/}} and DERI Pipes~\cite{Le-Phuoc2009}. The former approaches typically fail at integrating different data sources (e.g. \textit{list five cheapest offers among Amazon and eBay feeds}), whereas aggregation services are limited to filter results by brittle regex-based expressions (e.g. \textit{show only shops in New York}) and lack simple unit conversion (e.g. \textit{display all prices in euros}) (cf. \cite{Stolz2013:Hypertext}).

\section{Conclusions}

The presented on-line service aims to address the issue of Linked Open Data (LOD) content integration in Web pages. For this purpose, it generates RSS and Atom feeds from structured e-commerce data fetched from a SPARQL endpoint, thereby exploiting the excellent tool support for content syndication formats. The service allows for different levels of query building assistance, implements caching, incorporates geo-location data and RDFa annotations, and employs currency conversion at the endpoint. We consider our approach to be of similar use for other fields of Linked Open Data outside the narrow scope of e-commerce.

%We outlined the different search modes with different level of query building assistance. Then we explained how caching is implemented, presented the inclusion of geopositional information in data feeds, described the RDFa piggybacking approach in RSS and Atom to prevent proliferation of identifiers, and exemplified how currency conversion was implemented.

%Our approach has the following important implications:
%(1)~Site owners can add relevant, dynamic content to their Web pages, thereby attracting visitors and improving search engine ranking; 
%(2)~vendors gain additional visibility of their products for free, since items republished by virtue of site owners and bloggers link back to the shops where the products are actually sold;
%(3)~RDFa piggybacking in RSS and Atom further leverages the proliferation of product offers;
%(4)~repeated republishing of product offers opens up unprecedented market potential for partner programs. It allows for affiliate scenarios similar to what large marketplaces already provide. For instance, Amazon and eBay support the creation of customized feeds for registered affiliates, who can earn a commission for every purchase made through their personal affiliate links.

% an extension could be to use pubsubhubbub instead of caching and prefetching mechanism currently employed

%\section{References}

\bibliographystyle{splncs03}
\bibliography{gr2rss_demo}

\end{document}